\documentclass[%
 preprint,
 amsmath,amssymb,
 aps,
]{revtex4}

\usepackage{graphicx}
\usepackage{dcolumn}
\usepackage{bm}

\usepackage[utf8]{inputenc}
\usepackage[T1]{fontenc}
\usepackage{ulem}
\usepackage{color}

\begin{document}

\preprint{Comptes Rendus Acad\'emie des Sciences}

\title{Morphogenesis through elastic phase separation \\  in a pneumatic surface}
\thanks{A tribute to Yves Couder }%

\author{Emmanuel Si\'efert}
 \email{emmanuel.siefert@gmail.com}
\author{Beno\^it Roman}%
 \email{benoit.roman@espci.fr}
 \author{}%
\affiliation{%
PMMH, CNRS, ESPCI Paris, Universit\'e PSL, Sorbonne Universit\'e Universit\'e de Paris, F-75005, Paris, France
}%

\date{\today}

\begin{abstract}
We report a phenomenon of phase separation that relates in many aspects {to} Yves Couder's work: an inflatable architectured elastomer plate, expected to expand  homogeneously in its plane, buckles  instead widely out-of-plane into very complex shape when internal pressure is applied. We show that this morphogenetic pattern formation is due to a two-dimensional elastic phase separation, which induces incompatible {patchy} non-Euclidean reference metric.
\begin{description}
\item[Keywords] Pattern formation, elastic plate, morphogenesis
\end{description}
\end{abstract}

\maketitle


\section{Introduction}

Morphogenesis has been one of the subjects of interest of Yves Couder, with a strong contribution on how  geometry and mechanics play a role in plant spatial organisation of organs \cite{Douady92, Couder02}, and shape regulation~\cite{Corson09}. The study of instability in mechanics is another important string of works by Yves Couder \cite{Thome89, Boudaoud00}.
Here we report a new elastic phase-separation instability,  that leads to very impressive shape morphogenesis.

\begin{figure}[htbp]
  \centering
    \includegraphics[width= \textwidth]{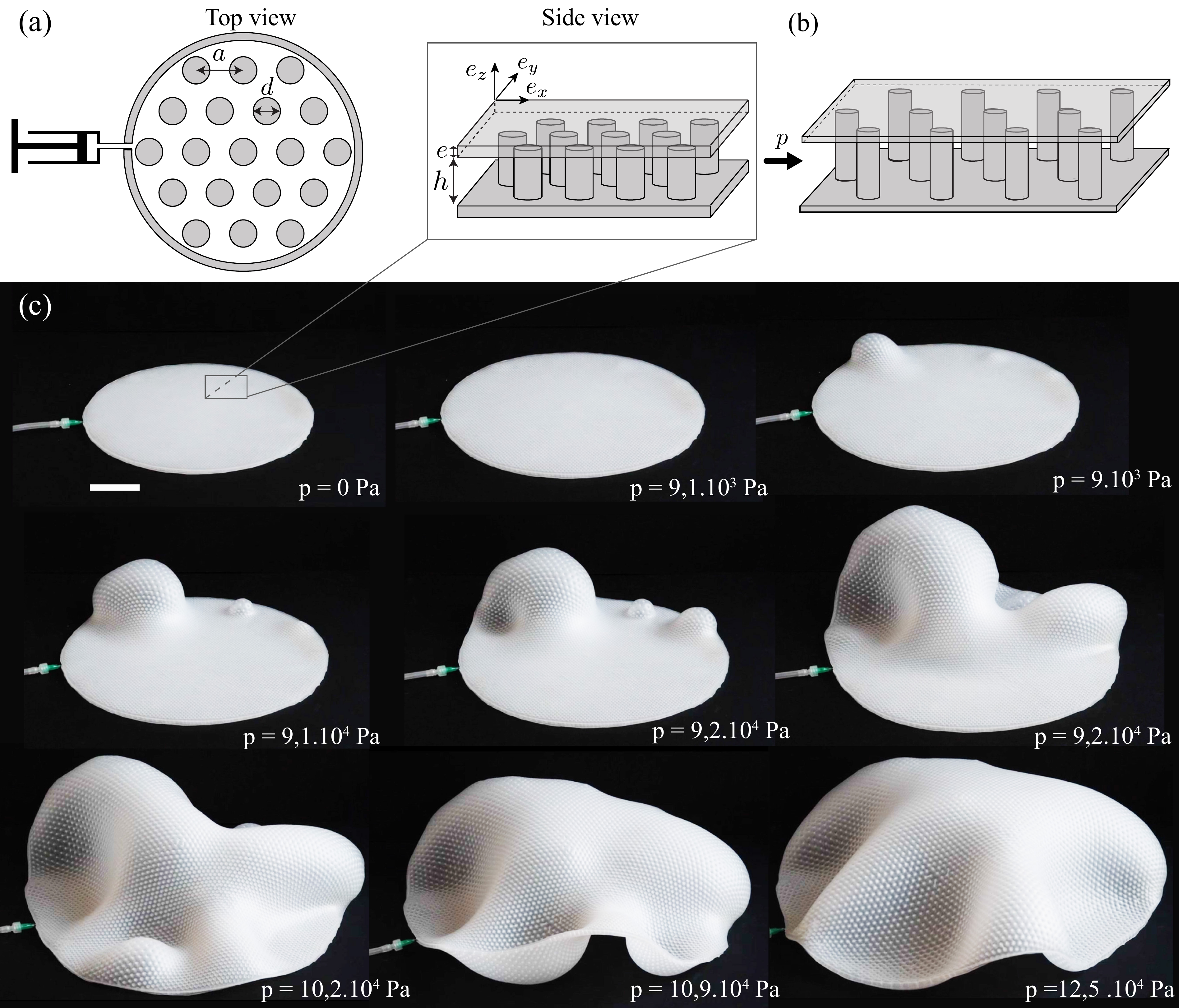}
      \caption{ Elastic phase separation. (a) (left) Schematic top view of the regular pillar lattice of pitch $a$, the diameter of the pillars being $d$.  {The boundary walls have a thickness of approx. $a/2$. The structure is connected to a pressure supply (e.g. a syringe).} (right) Schematic 3D representation of a portion of the structure, $e$ being the thickness of the top and bottom membranes, and $h$ the height of the pillars. (b) Upon inflation, the structure deforms, changing thus the actual $\psi$ and $\phi$ parameters, leading to geometric non-linearities. In the model proposed, the deformed pillars are approximated by cylinders, neglecting the boundary effect on the actual shape of the pillar, and accepting the deformation mismatch at the boundary pillar-membrane (the pillars shrink in the xy-plane, whereas the membrane is stretched).  {(c) Series of pictures of a large plate embedding a regular hexagonal lattice of pillars upon inflation (diameter of 25 cm, a=3 mm, d=2.4 mm, h=3 mm, e=1.2 mm). Bulges appear in the plate and then propagate through the whole structure, at a nearly constant pressure.  Scale bar: 5 cm.}}
      \label{fighexarest}
\end{figure}

We consider an architected elastomeric plate, with the following internal structure: a triangular network (spacing $a$) of cylindrical pillars  (height $h$ and diameter $d$) connects a top and bottom layer (thickness $e$), as in Fig.~\ref{fighexarest}. 
When a pressure $p$ is imposed in this internal chamber, which has a internal structure everywhere identical, we can expect a uniform response of the material, 
so that the plate should simply expand everywhere by the same factor. 
This is indeed the case when pressure is low (see first 2 pictures in Fig. \ref{fighexarest}(c)). But  very surprising shape changes are observed at one critical pressure (pressure is approximately constant in all pictures except the first one on Fig. \ref{fighexarest}(c)), where the structure locally buckles out of plane in several areas, whereas the rest remains flat. As more air is inflated inside the plate, the shape evolves as the bumps progressively merge though a complex series of shapes in a very homogeneous structure. 

Note that the homogeneously internal architecture presented here is different from the case of pneumatic shape-morphing elastomers, termed as {\it baromorphs} \cite{Siefert18}, where the geometry and orientation of internal cavities are spatially varied: when inflated, the resulting non-homogeneous expansion distorts the metric in a programmed non-Euclidean way, creates internal stresses in the plane and leads to the buckling of the structure. Here, in the contrary, the architecture is strictly periodic, isotropic and everywhere similar, so that a simple homogeneous planar expansion is expected. 

In this article we wish to elucidate this surprising "morphogenetic" instability. We start by a computation of the mechanical deformation of the inflated plate  if assumed to be homogeneous.  We next show that several states may be attained for the same imposed internal pressure and a phase separation instability is expected; the coexistence of the two phases (in strongly different stretch state and thus geometrically incompatible) explains the  complex shapes observed.

\section{Model and assumptions}
We start by a description of the plate deformation as a function of pressure of such an architectured plate (Fig. \ref{fighexarest}).  {Throughout this paper, the ratio $h/a$ is fixed equal to 1}. The geometry of the system is then defined by the two remaining relevant geometric parameters, $\Psi= h/(h+2e)$  the relative height of the pillars and $\Phi=\frac{\pi}{2\sqrt{3}}(\frac{d}{a})^2$ the in-plane pillar density (the corresponding airways density thus reads $1-\Phi$). 

For the sake of simplicity, we assume that the top and bottom membranes are subject to spatially homogeneous equibiaxial extension in the  $\mathbf{e}_x$ and $ \mathbf{e}_y$ directions (see Fig.\ref{fighexarest}(b)): the principal stretches are $ \lambda _{x}=\lambda _{y}=\lambda $. From incompressibility, $\lambda_x~\lambda_y~\lambda_z=1$. Hence  $\lambda_z=1/\lambda^2.$ Therefore, the first strain invariant, which quantifies the global amount of stretching of the molecular chains is 
\begin{equation}
   J_1 = \lambda_x^2+\lambda_y^2+\lambda_z^2-3 = 2~\lambda^2 + \frac{1}{\lambda^4}-3 
 \end{equation}
The material undergoes large strains in this experiment, and we use the phenomenological Gent model~\cite{gent96} that takes into account the material non-linearities (strain stiffening). The strain energy density function is designed such that it has a singularity when the first invariant $J_1$ reaches a limiting value $J_{m}$. Physically, it means that when the molecular chains are fully stretched, there is no possibility for further stretch: in practice, rupture arises before this theoretical limit.
According to Gent's model, we have the following constitutive law:
\begin{equation}
\underline{\underline{\sigma}}=-p_r\boldsymbol{I}+\frac{\mu J_m}{J_m-J_1} \boldsymbol{B}
\label{eq:gentconstitutive}
\end{equation}
where $p_r$ is a bulk pressure to be determined (through volume conservation), $ \boldsymbol{I}$ is the identity matrix, $\boldsymbol{B}$ the left Cauchy-Green deformation tensor and $\mu=E/2(1+\nu)$ is the shear modulus of the material.

The left Cauchy-Green deformation tensor in the top and bottom membranes can then be expressed as
\begin{equation}
   \boldsymbol{B} = \lambda^2~\mathbf{e}_x\otimes\mathbf{e}_x + \lambda^2~\mathbf{e}_y\otimes\mathbf{e}_y+ \frac{1}{\lambda^4}~\mathbf{e}_z\otimes\mathbf{e}_z 
 \end{equation}
Assuming that $\sigma_{zz}=0$ in the membrane, we get the following expression:
\begin{equation}
\sigma _{xx}=\sigma _{yy}=\left(\lambda ^{2}-{\frac {1}{\lambda ^{4}}}\right)\left({\frac {\mu J_{m}}{J_{m}-J_{1}}}\right).
\end{equation}
Moreover, we get through simple force balance, that
\begin{equation}
\sigma_{xx}=\sigma_{yy}=p\frac{\psi}{1-\psi}
\end{equation}
where 
\begin{equation}
\psi=\frac{\Psi\lambda^{p}}{\Psi\lambda^{p}+(1-\Psi)/\lambda^2}
\end{equation}
 is the actual relative height of the pillars in the deformed configuration. Pillars are indeed stretched by an amount $\lambda^p$ and the top and bottom membranes are thinner by an amount $1/\lambda^2$ due to material incompressibility.
 It is thus critical to compute the deformation $\lambda^{p}$ of the pillar in the vertical direction. Deformation and stresses in the pillars are denoted with the superscript $^p$. Pillars are compressed in the horizontal plane by the pressure inside the chamber: hence, $\sigma^p_{xx}=\sigma^p_{yy}=-p$. In the vertical direction, balancing the forces perpendicular to a cut in the $xy$-plane, yields:
\begin{equation}
\sigma^{p}_{zz}=p\dfrac{1-\phi}{\phi}
\end{equation}
where
\begin{equation}
\phi=\cfrac{\Phi}{\lambda^2\lambda^p}
\end{equation}
 is the actual pillar density in the deformed configuration. The pillar cross-section is indeed reduced whereas the elementary size of the lattice increases, both effects leading to a  reduction of the pillar density.
The symmetries of the pillars yield $ \lambda^p _{z}=\lambda^p ,~\lambda^p_{x}=\lambda^p_{y}$. From incompressibility, we get $ (\lambda _{x}^{p})^2=(\lambda _{y}^{p})^2=1/\lambda^p $. Therefore,
\begin{equation}
J_{1}=(\lambda _{x}^{p})^2+(\lambda _{y}^{p})^2+ (\lambda _{z}^{p})^2-3=(\lambda ^{p})^2 +{\cfrac {2}{\lambda^p }}-3
\end{equation}
 The left Cauchy-Green deformation tensor can then be expressed in the pillar as

\begin{equation}
   \boldsymbol{B^p} = \cfrac{1}{\lambda^p}~(\mathbf{e}_x\otimes\mathbf{e}_x+\mathbf{e}_y\otimes\mathbf{e}_y)+ (\lambda^{p})^2~\mathbf{e}_z\otimes\mathbf{e}_z
\end{equation}
 Hence, we have
\begin{equation}
\sigma^p _{zz}=-p_r+{\cfrac {(\lambda ^{p})^2\mu J_{m}}{J_{m}-J_{1}}}~;~~\sigma ^p_{xx}=-p_r+{\cfrac {\mu J_{m}}{\lambda^p (J_{m}-J_{1})}}=\sigma^p _{yy}
\end{equation}
where $p_r$ is a bulk pressure coming from the incompressibility that we shall now determine. As $\sigma ^p_{xx}=\sigma^p _{yy}=-p$, we have
\begin{equation}
 p_r= p+{\cfrac {\mu J_{m}}{\lambda^p (J_{m}-J_{1}})}
\end{equation}
Therefore,
\begin{equation}
\sigma ^p_{zz}=-p+\left((\lambda ^{p})^2-{\frac {1}{\lambda^p }}\right)\left({\frac {\mu J_{m}}{J_{m}-J_{1}}}\right)
\end{equation}
The actual in-plane pillar density $\phi$ in the deformed state reads thus $\phi=\Phi/(\lambda^2\lambda^p)$.

To summarize, we propose a very simple model, in which the physical link of the pillars on the membrane (i.e. boundary conditions on the pillar and actual stiffness of the membrane due to the presence of pillars) is completely overseen. This model is believed to be more accurate when both $\Psi\to 1$ and $\Phi\to 0$, that is when the pillars are slender structures and do not affect much the membranes stretching (see Figure \ref{fighexarest}(b)).
\begin{figure}
  \centering
    \includegraphics[width=0.6\textwidth]{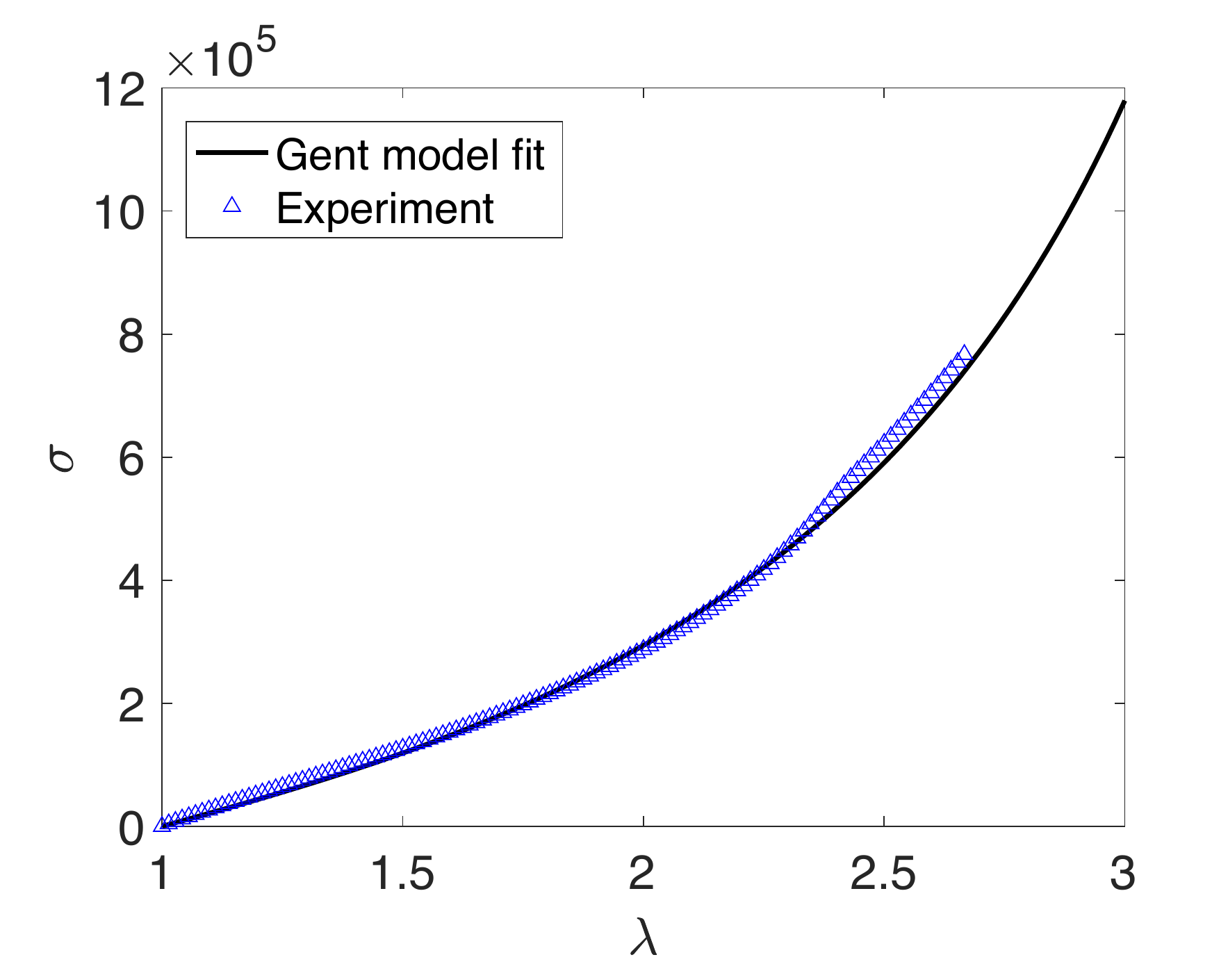}
      \caption{ Fit of a tensile test on a ribbon made of Elite Double 8 from Zhermack with the incompressible Gent model: the two parameters, namely the shear modulus $\mu$ and the limited value of the first invariant $J_m$ are found to best fit the experimental curve for the values of $7.2.10^4$ Pa and 14.2 respectively. }
      \label{gentfit}
\end{figure}

\noindent
The system to be solved is thus the following:
\begin{equation}
\left
\lbrace
\begin{array}{cc}
\sigma _{xx}=p\cfrac{\psi}{1-\psi}=\left(\lambda ^{2}-{\cfrac {1}{\lambda ^{4}}}\right)\left({\cfrac {\mu J_{m}}{J_{m}-2~\lambda^2 - \cfrac{1}{\lambda^4} +3}}\right)\\\\
\sigma ^p_{zz}=p\cfrac{1-\phi}{\phi}=-p+\left((\lambda ^{p})^2-{\cfrac {1}{\lambda^p }}\right)\left({\cfrac {\mu J_{m}}{J_{m}-(\lambda ^{p})^2-{\cfrac {2}{\lambda^p }}+3}}\right)\\\\
\phi=\cfrac{\Phi}{\lambda^2\lambda^p}\\\\
\psi=\cfrac{\Psi\lambda^{p}}{\Psi\lambda^{p}+(1-\Psi)/\lambda^2}
\end{array}
\right.
\label{eq2Dbulge}
\end{equation}
\noindent
where the unknowns are $\lambda$ and $\lambda^p$, $\mu$ and $J_m$ are material properties, $\Psi$ and $\Phi$ are geometric parameters of the structures and $p$ the applied pressure.

\section{Experimental realisation and quantitative comparison with the model}

A tensile test is first performed at low speed to measure the characteristics of the silicone elastomer as shown in Fig. \ref{gentfit} for the case of Elite Double 8 from Zhermack (see Table \ref{tab:gent} for the typical values measured during this thesis). Fitting the experimental stress-deformation curve with the  theoretical constitutive law of Gent model for incompressible hyperelastic materials, one infers the shear modulus of the material $\mu$ and the limiting value of the first invariant $J_m$ for each elastomer.
\begin{table}
\centering
\begin{tabular}{|l|r|r|}
  \hline
  Material & shear modulus $\mu$ & First invariant limit $J_m$ \\
  \hline
  Ecoflex 0050 & $3,0.10^4$ Pa& 23 \\
  Elite Double 8 & $7,2.10^4$ Pa& 14.2 \\
DragonSkin 10 & $15.10^4$ Pa& 18 \\
DragonSkin 20& $21.10^4 $ Pa& 7.6 \\
  \hline
\end{tabular}
\caption{Material properties deduced from a simple traction test by fitting the curve with the expected response of a Gent hyperelastic model. }
\label{tab:gent}
\end{table}

The plates with internal pillar structures are made of silicone elastomers (see table \ref{tab:gent}) in a 3 steps process.
Equal quantities of catalyst and base liquids are mixed and placed
in a vacuum chamber to remove trapped air bubble. Two sheets of thickness $e$ are spread on a flat surface. After curing, a circular 3D-printed annulus (thickness $h$, inner radius $R$) is placed on top of one sheet and a new mixture is poured inside the annulus. A 3D-printed  perforated plate (radius $R-d/2$, thickness $h$, holes of diameter $d$ forming a triangular lattice of pitch $a$) is then placed inside the annulus and pushed through the liquid mixture to reach the bottom elastomeric sheet. Exceeding liquid above the plate is scraped with a ruler. After curing, the structure (now comprising the sheet and the the pillars) is  removed from the perforated plate and is finally ‘glued’ to the remaining flat sheet using a thin layer of uncured mixture of the same material.

\begin{figure}
  \centering
    \includegraphics[width=\textwidth]{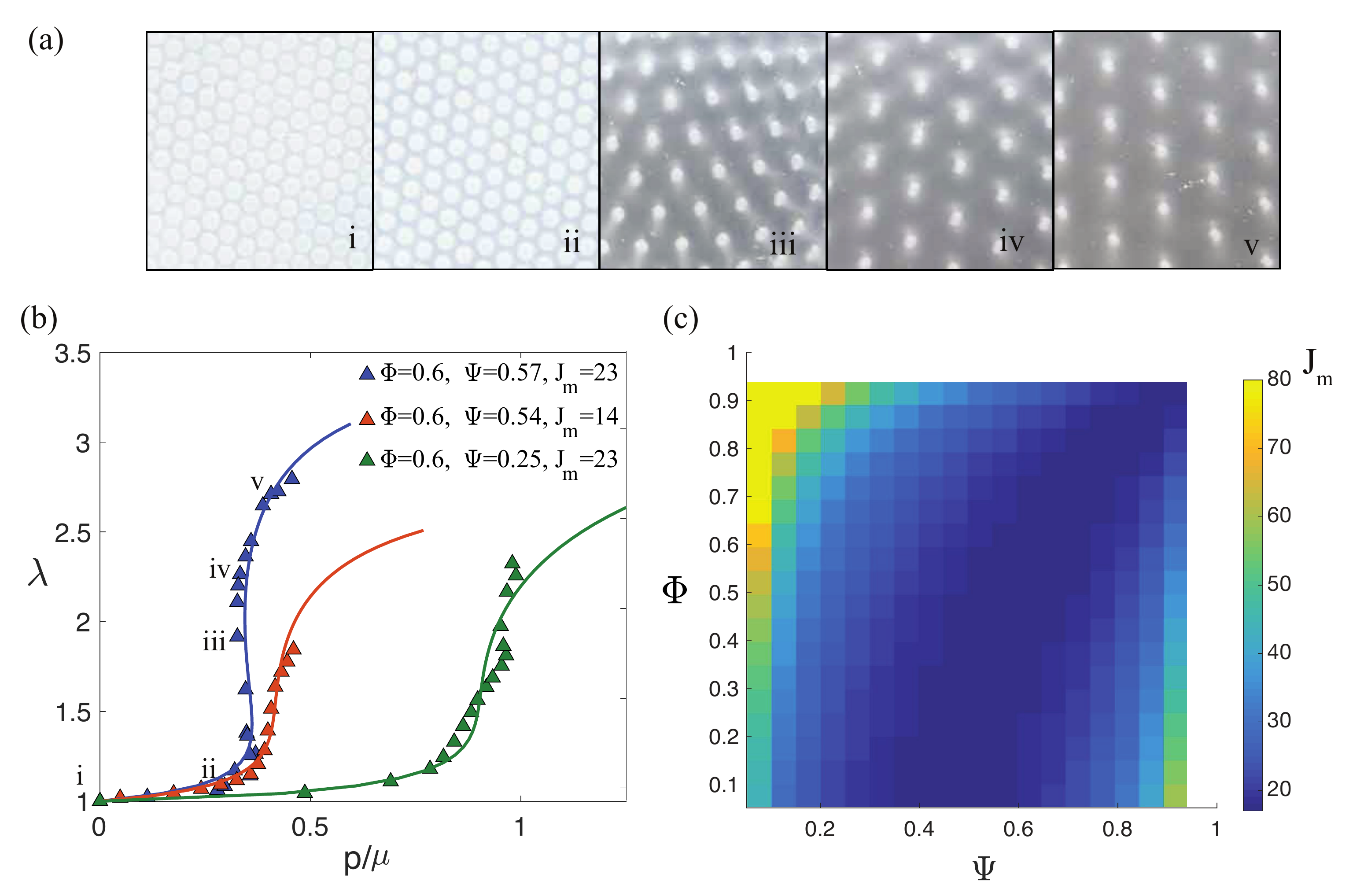}
      \caption{
      (a) Series of pictures of a portion of a plate for various imposed volumes. (b) Deformation response of the plate as a function of the dimensionless applied
pressure $p/\mu$ for various values of $\Psi, \Phi$ and $J_m$. Triangles correspond to experimental
measurements, solid lines to the theoretical response. Snapshots in (a) correspond to some data points on the blue curve. The deformation may be
easily measured by following the evolution of the average distance between pillar centres.
(c) Minimum value of $J_m$ (color scale) in order to have the superelastic instability as a
function of the geometrical parameters $\Psi$ and $\Phi$."
       }
      \label{figsuper}
\end{figure}

 {The experimental deformation in the pressurized structure is measured by taking top view pictures at various stages, as shown in Fig. \ref{figsuper}(a). In regions where the in plane expansion is homogeneous, the evolution of the local mean distance between two pillars is tracked to extract the stretching factor $\lambda$ as a function of pressure (Fig. \ref{figsuper}(b)). The volume, rather than the pressure, of air injected in the structure is controlled in order to reach as many equilibrium states as possible during both inflation and deflation.}
Solving the system of equations (\ref{eq2Dbulge}) with Matlab, the computed theory obtained is in good quantitative agreement with the experimental data points without any fitting parameter (see Fig. \ref{figsuper}(b)). Both the linear response at small pressure and the strong non-linearities at larger pressure are well predicted by the model. 
We obtain, for long enough chains ({\it i.e.} for large $J_m$), an S-curve.
It means that two phases may coexist for one given pressure, one stretched, the other barely strained. 

\begin{figure}[htbp]
  \centering
    \includegraphics[width=0.6\textwidth]{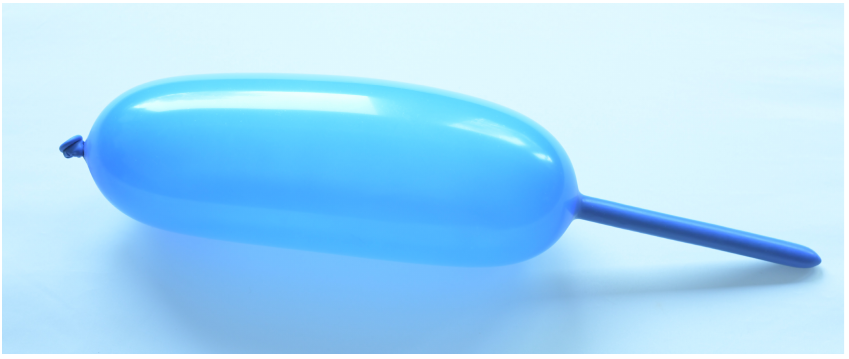}
      \caption{An inflated tubular balloon is an example of phase separation in a 1D system: a portion with very large expansion of the diameter (on the left of the balloon) coexists with a portion in a phase with rather small deformation (right side of the balloon).  }
      \label{figballon1D}
\end{figure}

When the mechanical response of a system (assumed homogeneous) includes a regime where strain decreases with load, and later a subsequent stiffening (i.e. S shaped curved), a phenomenon analogous to a phase separation occurs.
In such cases, a Maxwell construction~\cite{Chater84} leads to a constant load for which two different state coexist in a spatially extended system. Such phase transition are well known in mechanics, and  rather termed "localization", "propagative instability" or "coexistent phases".  
Examples are the hysteretic buckling of the tape-spring measurer~\cite{Seffen99}, the collapse of depressurized cylinders~\cite{Kyriakides93}, the rippling of multi-walled carbon nanotubes~\cite{Arroyo08}, the necking instability in plastic bar~\cite{Ericksen75, Audoly19} or the plateau in the stress-strain response for elastic foams~\cite{Gioia01}.
But maybe the simplest and more common example is the inflation of a cylindrical hyperelastic ballon~\cite{Chater84}, which presents regions with very large diameter expansion coexisting with regions hardly stretched (see Fig.~\ref{figballon1D}).

Our system depends on two geometric parameters, namely $\Phi$ and $\Psi$, that may impact this bulge instability. 
In Figure \ref{figsuper}(c), we show for each possible pair ($\Psi,\Phi$), the minimum limiting first invariant value $J_m$ in order to get this instability. It appears that the in-plane relative stiffness, measured by $\Psi$, and the vertical stiffness, measured by $\Phi$, must typically have the same value in order to get the instability at low $J_m$.

Because they are one-dimensional objects, roots that grow non-uniformly, bars under localized plastic extension~\cite{Ericksen75}, or cylindrical balloon~\cite{Chater84} (figure~\ref{figballon1D}), the resulting inhomogeneous axial stretch does not produce geometrical incompatibilities. They
are free to deform according to the growth distribution and therefore do not develop internal stresses. Conversely, non-uniform growth of two-dimensional sheets can be geometrically incompatible~\cite{Klein07, Kim12}, leading to the accumulation of stresses within the sheet. For example, if one portion of the sheet grows more rapidly than its surrounding, it may buckle out of plane, as studied by Bense {\it et al.} \cite{bense17} in the case of an expanding patch of dielectric elastomer in a passive elastomer sheet.
In the cylindrical balloon configuration, the apparition of a bulge, which is much more stretched than the rest of the balloon, is not affected by the barely stretched rest of the ballon. There is a smooth transition between the two phases, but no internal stresses build up in the balloon, as a nearly 1D structure.
In the 2D sheet, conversely, the apparition of a bulge is geometrically incompatible with its barely stretched surrounding. The structure thus locally buckles out of plane, as shown in Fig. \ref{fighexarest}(c), leading to a complex topology of the very simple and initially regular internal structure. The two phases corresponding to the same pressure, the structure picks the proportion of the highly stretch phase depending on the air volume inserted in the structure.  

Although the target strain-pressure curve and the subsequent bulge apparition are well understood and captured by our minimal model, the induced 3D shapes are beyond the scope of this article. They involve the theory of non Euclidean incompatible plates at finite deformation with a specific superelastic response.

\begin{figure}[htbp]
  \centering
    \includegraphics[width=0.8\textwidth]{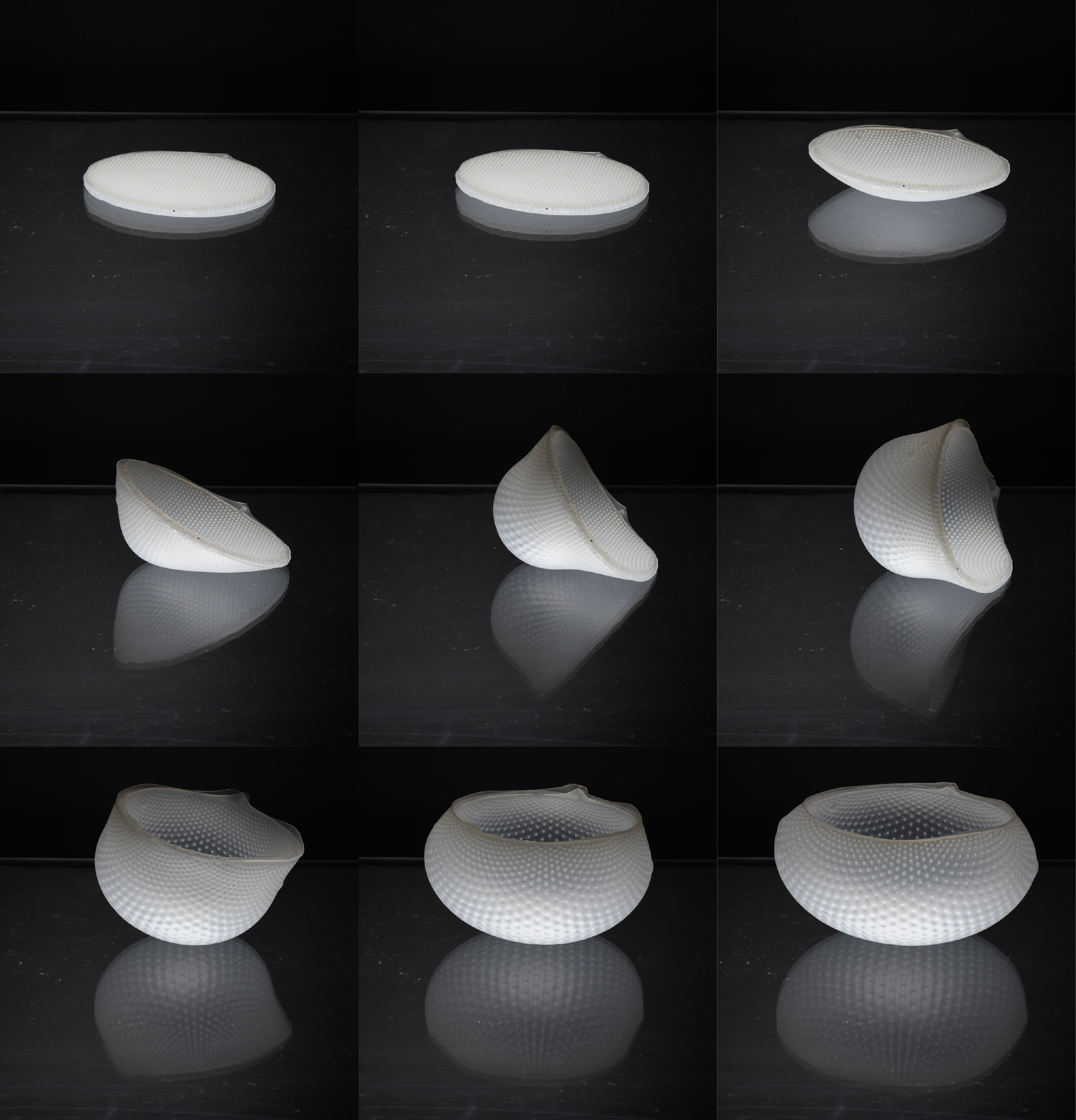}
      \caption{{Pneumatic inflation of an architected elastomeric disk  with the same internal structure as in Figure~\ref{fighexarest}, but with a smaller external radius $R$. Only one bulge appears here and, together with the constraint imposed by the outer walls, the structure shapes into a bowl. } }
      \label{fighexabowl}
\end{figure}

When the plate is smaller (but the thickness stays the same), its radius $R$  gets closer to the typical boundary layer width between the two phases. The structure can thus encompass only one bulge and the shape is more controlled; the presence of an elastomeric wall around the structures frustrates the in-plane extension of the structure and a well-defined bowl shape is obtained (Fig. \ref{fighexabowl}).  

\section{Conclusion}
We have reported in this article a surprising shape morphogenesis from an uniform object, expected to expand uniformly. In contrast with the shape-morphing triggered by programmed inhomogeneous (incompatible) in-plane expansion, here the internal structure of the plate is identical everywhere, but phase separation induces area with very large expansion to co-exist with regions with very little expansion. 
This is the 2D (plate) equivalent of the tubular balloon phase separation. because it is 2D, geometrical incompatibilities lead to out-of plane buckling.

We have only described here the reason for the instability and many questions remain: can we predict the  shapes obtained? What sets the size of the boundary layer between the phases? what is the mechanical response of such phase-separated shells?

We however had the opportunity to show and discuss this phenomenon with Yves Couder, a few months before he passed away. Despite his physical handicap at that time, Yves enjoyed the experiments, and the discussion then drifted around the phyllotaxy of new algies that appeared in Bretagne; the stress concentration experienced when your partly non-responsive body has to be moved by other people; the work by an artist friend who performs cuts in a cube of mattress foam, which only takes its programmed shape once reversed inside-out… 
This journey through apparently disconnected subjects was in fact about the relation between forces, geometry and shape.
And indeed, for somebody very interested also in etymology as Yves Couder, it is not a coincidence that «comprendre» («to understand» in French) originates from the latin «cum» (together) and «prehendere» (take/bring): for him, to understand is really to «bring together», draw analogy and unify apparently disconnected phenomena.

\bibliography{CRAS_Siefert_Roman.bib}

\end{document}